\documentclass[a4paper,10pt,english]{article}
\usepackage[utf8]{inputenc}
\usepackage[T1]{fontenc}
\usepackage{babel}
\usepackage{times}
\usepackage[Bjarne]{fncychap}
\usepackage{longtable}
\usepackage{sphinx}

\title{pydelay -- a python tool for solving delay differential equations\\v0.1.1}
\date{November 09, 2009}
\release{0.1.1}
\author{V. Flunkert \and E. Schöll}

\renewcommand{\releasename}{Release}
\makeindex

\begin{document}

\maketitle
\centerline{Contact: flunkert@itp.tu-berlin.de}
\hypertarget{--doc-index}{}

\section{Introduction}

pydelay is a program which translates a system of delay differential equations
(DDEs) into simulation C-code and compiles and runs the code (using scipy
weave). This way it is easy to quickly implement a system of DDEs but you still
have the speed of C. The Homepage can be found here:
\begin{quote}

\href{http://pydelay.sourceforge.net/}{http://pydelay.sourceforge.net/}
\end{quote}

It is largely inspired by \href{http://www.cam.cornell.edu/~rclewley/cgi-bin/moin.cgi/}{PyDSTool}.

The algorithm used is based on the Bogacki-Shampine method \footnote{
Bogacki, P. and Shampine, L. F., A 3(2) pair of Runge - Kutta formulas, Applied Mathematics Letters 2, 4, 321 ISSN 0893-9659, (1989).
}
which is also implemented in \emph{Matlab's dde23} \footnote{
Shampine, L. F. and Thompson, S., Solving DDEs in Matlab, Appl. Num. Math. 37, 4, 441 ( 2001)
}.

We also want to mention \href{http://users.ox.ac.uk/~clme1073/python/PyDDE/}{PyDDE}
-- a different python program for solving DDEs.

\textbf{License}

pydelay is licensed under the MIT License.

\subsection{Installation and requirements}

\textbf{Unix:}

You need \href{http://www.python.org/}{python}, \href{http://numpy.scipy.org/}{numpy and scipy} and the \emph{gcc}-compiler.
To plot the solutions and run the examples you also need \href{http://matplotlib.sourceforge.net/index.html}{matplotlib}.

To install pydelay grab the latest \emph{tar.gz} from the website and install the package in the usual way:

\begin{Verbatim}[commandchars=@\[\]]
cd pydelay-@$version
python setup.py install
\end{Verbatim}

When the package is installed, you can get some info about the functions and the usage with:

\begin{Verbatim}[commandchars=@\[\]]
pydoc pydelay
\end{Verbatim}

\textbf{Windows:}

The solver has not been tested on a \emph{windows} machine. It could perhaps work
under \href{http://www.cygwin.com/}{cygwin}.

\subsection{An example}

The following example shows the basic usage. It solves the Mackey-Glass
equations \footnote{
Mackey, M. C. and Glass, L. (1977). Pathological physiological conditions resulting from instabilities in physiological control system. Science, 197(4300):287-289.
} for initial conditions which lead to a periodic orbit (see \footnote{
\href{http://www.scholarpedia.org/article/Mackey-Glass\_equation}{http://www.scholarpedia.org/article/Mackey-Glass\_equation}
}
for this example).

\begin{Verbatim}[commandchars=@\[\]]
@PYGaE[@# import pydelay and numpy and pylab]
@PYGal[import] @PYGaW[numpy] @PYGal[as] @PYGaW[np]
@PYGal[import] @PYGaW[pylab] @PYGal[as] @PYGaW[pl]
@PYGal[from] @PYGaW[pydelay] @PYGal[import] dde23

@PYGaE[@# define the equations]
eqns @PYGbe[=] {
    @PYGaB[']@PYGaB[x]@PYGaB['] : @PYGaB[']@PYGaB[0.25 * x(t-tau) / (1.0 + pow(x(t-tau),p)) -0.1*x]@PYGaB[']
    }

@PYGaE[@#define the parameters]
params @PYGbe[=] {
    @PYGaB[']@PYGaB[tau]@PYGaB[']: @PYGaw[15],
    @PYGaB[']@PYGaB[p]@PYGaB[']  : @PYGaw[10]
    }

@PYGaE[@# Initialise the solver]
dde @PYGbe[=] dde23(eqns@PYGbe[=]eqns, params@PYGbe[=]params)

@PYGaE[@# set the simulation parameters]
@PYGaE[@# (solve from t=0 to t=1000 and limit the maximum step size to 1.0)]
dde@PYGbe[.]set@_sim@_params(tfinal@PYGbe[=]@PYGaw[1000], dtmax@PYGbe[=]@PYGaw[1.0])

@PYGaE[@# set the history of to the constant function 0.5 (using a python lambda function)]
histfunc @PYGbe[=] {
    @PYGaB[']@PYGaB[x]@PYGaB[']: @PYGay[lambda] t: @PYGaw[0.5]
    }
dde@PYGbe[.]hist@_from@_funcs(histfunc, @PYGaw[51])

@PYGaE[@# run the simulator]
dde@PYGbe[.]run()

@PYGaE[@# Make a plot of x(t) vs x(t-tau):]
@PYGaE[@# Sample the solution twice with a stepsize of dt=0.1:]

@PYGaE[@# once in the interval @PYGZlb[]515, 1000@PYGZrb[]]
sol1 @PYGbe[=] dde@PYGbe[.]sample(@PYGaw[515], @PYGaw[1000], @PYGaw[0.1])
x1 @PYGbe[=] sol1@PYGZlb[]@PYGaB[']@PYGaB[x]@PYGaB[']@PYGZrb[]

@PYGaE[@# and once between @PYGZlb[]500, 1000-15@PYGZrb[]]
sol2 @PYGbe[=] dde@PYGbe[.]sample(@PYGaw[500], @PYGaw[1000]@PYGbe[-]@PYGaw[15], @PYGaw[0.1])
x2 @PYGbe[=] sol2@PYGZlb[]@PYGaB[']@PYGaB[x]@PYGaB[']@PYGZrb[]

pl@PYGbe[.]plot(x1, x2)
pl@PYGbe[.]xlabel(@PYGaB[']@PYGaB[@$x(t)@$]@PYGaB['])
pl@PYGbe[.]ylabel(@PYGaB[']@PYGaB[@$x(t - 15)@$]@PYGaB['])
pl@PYGbe[.]show()
\end{Verbatim}

\includegraphics{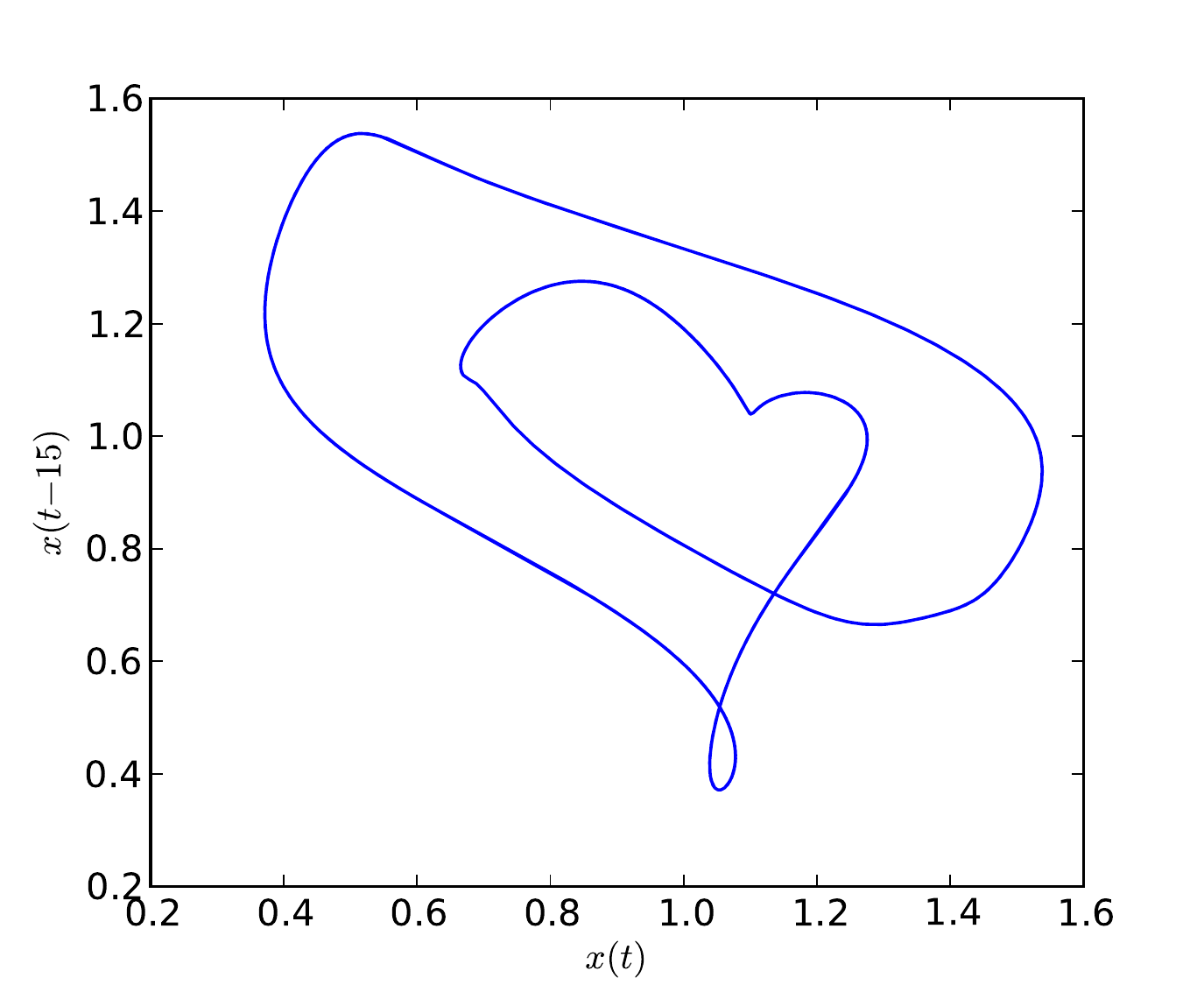}

\section{Usage}

\subsection{Defining the equations, delays and parameters}

Equations are defined using a python dictionary.
The keys are the variable names and the entry is the right hand side of the
differential equation.
The string defining the equation has to be a valid C expression, i.e.,
use \code{pow(a,b)} instead of \code{a**b} etc.

Delays are written as \code{(t-delay)}, where \code{delay} can be
some expression involving parameters and numbers but not (yet) involving
the time \code{t} or the dynamic variables:

\begin{Verbatim}[commandchars=@\[\]]
eqns @PYGbe[=] {
    @PYGaB[']@PYGaB[y1]@PYGaB[']: @PYGaB[']@PYGaB[- y1 * y2(t-tau) + y2(t-1.0)]@PYGaB['],
    @PYGaB[']@PYGaB[y2]@PYGaB[']: @PYGaB[']@PYGaB[a * y1 * y2(t-2*tau) - y2]@PYGaB['],
    @PYGaB[']@PYGaB[y3]@PYGaB[']: @PYGaB[']@PYGaB[y2 - y2(t-(tau+1))]@PYGaB[']
  }
\end{Verbatim}

Complex variables can be defined by adding \code{':c'} or
\code{':C'} in the eqn-dictionary.
The imaginary unit can be used through \code{'ii'} in the equations:

\begin{Verbatim}[commandchars=@\[\]]
eqns @PYGbe[=] {
    @PYGaB[']@PYGaB[z:c]@PYGaB[']: @PYGaB[']@PYGaB[(la + ii*w0 + g*pow(abs(z),2) )*z + b*(z(t-tau) - z(t))]@PYGaB['],
}
\end{Verbatim}

Parameters are defined in a separate dictionary where the keys are
the parameter names, i.e.,:

\begin{Verbatim}[commandchars=@\[\]]
params @PYGbe[=] {
    @PYGaB[']@PYGaB[a]@PYGaB[']  : @PYGaw[0.2],
    @PYGaB[']@PYGaB[tau]@PYGaB[']: @PYGaw[1.0]
}
\end{Verbatim}

\subsection{Setting the history}

The history of the variables is stored in the dictionary \code{dde23.hist}.
The keys are the variable names and there is an additional key \code{'t'} for
the time array of the history.

There is a second dictionary \code{dde23.Vhist}
where the time derivatives of the history is stored (this is needed for the
solver). When the solver is initialized, i.e.,:

\begin{Verbatim}[commandchars=@\[\]]
dde @PYGbe[=] dde23(eqns, params)
\end{Verbatim}

the history of all variables (defined in \code{eqns}) is initialized to an
array of length \code{nn=101} filled with zeros. The time array is evenly
spaced in the interval \code{{[}-maxdelay, 0{]}}.

It is possible to manipulate these arrays directly, however this is not
recommended since one easily ends up with an ill-defined history resulting for
example in segfaults or false results.

Instead use the following methods to set the history.
\index{hist\_from\_funcs() (pydelay.dde23 method)}

\hypertarget{pydelay.dde23.hist_from_funcs}{}\begin{methoddesc}[dde23]{hist\_from\_funcs}{dic, nn=101}
Initialise the histories with the functions stored in the dictionary \emph{dic}.
The keys are the variable names.  The function will be called as \code{f(t)} 
for \code{t} in \code{{[}-maxdelay, 0{]}} on \emph{nn} samples in the interval.

This function provides the simplest way to set the history.
It is often convenient to use python \code{lambda} functions for \code{f}.
This way you can define the history function in place.

If any variable names are missing in the dictionaries, the history of these
variables is set to zero and a warning is printed. If the dictionary contains 
keys not matching any variables these entries are ignored and a warning is 
printed.

Example: Initialise the history of the variables \code{x} and \code{y} with 
\code{cos} and \code{sin} functions using a finer sampling resolution:

\begin{Verbatim}[commandchars=@\[\]]
@PYGal[from] @PYGaW[math] @PYGal[import] sin, cos

histdic @PYGbe[=] {
    @PYGaB[']@PYGaB[x]@PYGaB[']: @PYGay[lambda] t: cos(@PYGaw[0.2]@PYGbe[*]t),
    @PYGaB[']@PYGaB[y]@PYGaB[']: @PYGay[lambda] t: sin(@PYGaw[0.2]@PYGbe[*]t)
}

dde@PYGbe[.]hist@_from@_funcs(histdic, @PYGaw[500])
\end{Verbatim}
\end{methoddesc}
\index{hist\_from\_arrays() (pydelay.dde23 method)}

\hypertarget{pydelay.dde23.hist_from_arrays}{}\begin{methoddesc}[dde23]{hist\_from\_arrays}{dic, useend=True}
Initialise the history using a dictionary of arrays with variable names as keys.  
Additionally a time array can be given corresponding to the key \code{t}. 
All arrays in \emph{dic} have to have the same lengths.

If an array for \code{t} is given the history is interpreted as points 
\code{(t,var)}. Otherwise the arrays will be evenly spaced out over the interval 
\code{{[}-maxdelay, 0{]}}.

If useend is True the time array is shifted such that the end time is
zero. This is useful if you want to use the result of a previous simulation 
as the history.

If any variable names are missing in the dictionaries, the history of these
variables is set to zero and a warning is printed. 
If the dictionary contains keys not matching any variables (or \code{'t'}) these
entries are ignored and a warning is printed.

Example::

\begin{Verbatim}[commandchars=@\[\]]
t @PYGbe[=] numpy@PYGbe[.]linspace(@PYGaw[0], @PYGaw[1], @PYGaw[500])
x @PYGbe[=] numpy@PYGbe[.]cos(@PYGaw[0.2]@PYGbe[*]t)
y @PYGbe[=] numpy@PYGbe[.]sin(@PYGaw[0.2]@PYGbe[*]t)

histdic @PYGbe[=] {
    @PYGaB[']@PYGaB[t]@PYGaB[']: t,
    @PYGaB[']@PYGaB[x]@PYGaB[']: x,
    @PYGaB[']@PYGaB[y]@PYGaB[']: y
}
dde@PYGbe[.]hist@_from@_arrays(histdic)
\end{Verbatim}
\end{methoddesc}

Note that the previously used methods \code{hist\_from\_dict}, \code{hist\_from\_array}
and \code{hist\_from\_func} (the last two without \code{s}) have been removed, since it
was too easy to make mistakes with them.

\subsection{The solution}

After the solver has run, the solution (including the history) is stored
in the dictionary \code{dde23.sol}. The keys are again the variable names
and the time \code{'t'}. Since the solver uses an adaptive step size method,
the solution is not sampled at regular times.

To sample the solutions at regular (or other custom spaced) times there
are two functions.
\index{sample() (pydelay.dde23 method)}

\hypertarget{pydelay.dde23.sample}{}\begin{methoddesc}[dde23]{sample}{tstart=None, tfinal=None, dt=None}
Sample the solution with \emph{dt} steps between \emph{tstart} and \emph{tfinal}.
\begin{description}
\item[\emph{tstart}, \emph{tfinal}] \leavevmode
Start and end value of the interval to sample.
If nothing is specified \emph{tstart} is set to zero
and \emph{tfinal} is set to the simulation end time.

\item[\emph{dt} ] \leavevmode
Sampling size used. If nothing is specified a reasonable 
value is calculated.

\end{description}

Returns a dictionary with the sampled arrays. The keys are the 
variable names. The key \code{'t'} corresponds to the sampling times.
\end{methoddesc}
\index{sol\_spl() (pydelay.dde23 method)}

\hypertarget{pydelay.dde23.sol_spl}{}\begin{methoddesc}[dde23]{sol\_spl}{t}
Sample the solutions at times \emph{t}.
\begin{description}
\item[\emph{t}] \leavevmode
Array of time points on which to sample the solution.

\end{description}

Returns a dictionary with the sampled arrays. The keys are the 
variable names. The key \code{'t'} corresponds to the sampling times.
\end{methoddesc}

These functions use a cubic spline interpolation of the solution data.

\subsection{Noise}

Noise can be included in the simulations. Note however, that the method used is
quite crude (an Euler method will be added which is better suited for noise
dominated dynamics). The deterministic terms are calculated with the usual
Runge-Kutta method and then the noise term is added with the proper scaling of
\code{\$\textbackslash{}sqrt\{dt\}\$} at the final step. To get accurate results one should use small
time steps, i.e., \code{dtmax} should be set small enough.

The noise is defined in a separate dictionary. The function \code{gwn()} can
be accessed in the noise string and is a Gaussian white noise term of unit
variance. The following code specifies an Ornstein-Uhlenbeck process.:

\begin{Verbatim}[commandchars=@\[\]]
eqns @PYGbe[=] { @PYGaB[']@PYGaB[x]@PYGaB[']: @PYGaB[']@PYGaB[-x]@PYGaB['] }
noise @PYGbe[=] { @PYGaB[']@PYGaB[x]@PYGaB[']: @PYGaB[']@PYGaB[D * gwn()]@PYGaB[']}
params @PYGbe[=] { @PYGaB[']@PYGaB[D]@PYGaB[']: @PYGaw[0.00001] }

dde @PYGbe[=] dde23(eqns@PYGbe[=]eqns, params@PYGbe[=]params, noise@PYGbe[=]noise)
\end{Verbatim}

You can also use noise terms of other forms by specifying an appropriate
C-function (see the section on custom C-code).

\subsection{Custom C-code}

You can access custom C-functions in your equations by
adding the definition as \code{supportcode} for the solver.
In the following example a function \code{f(w,t)} is defined through
C-code and accessed in the eqn string.:

\begin{Verbatim}[commandchars=@\[\]]
@PYGaE[@# define the eqn f is the C-function defined below]
eqns @PYGbe[=] { @PYGaB[']@PYGaB[x]@PYGaB[']: @PYGaB[']@PYGaB[- x + k*x(t-tau) + A*f(w,t)]@PYGaB['] }
params @PYGbe[=] {
    @PYGaB[']@PYGaB[k]@PYGaB[']  : @PYGaw[0.1],
    @PYGaB[']@PYGaB[w]@PYGaB[']  : @PYGaw[2.0],
    @PYGaB[']@PYGaB[A]@PYGaB[']  : @PYGaw[0.5],
    @PYGaB[']@PYGaB[tau]@PYGaB[']: @PYGaw[10.0]
}

mycode @PYGbe[=] @PYGaB["""]
@PYGaB[double f(double t, double w) {]
@PYGaB[    return sin(w * t);]
@PYGaB[}]
@PYGaB["""]

dde @PYGbe[=] dde23(eqns@PYGbe[=]eqns, params@PYGbe[=]params, supportcode@PYGbe[=]mycode)
\end{Verbatim}

When defining custom code you have to be careful with the types.
The type of complex variables in the C-code is \code{Complex}.
Note in the above example that \code{w} has to be given as an input to
the function, because the parameters can only be accessed from the eqns string
and not inside the supportcode. (Should this be changed?)

Using custom C-code is often useful for switching terms on and off.
For example the Heaviside function may be defined and used as follows.:

\begin{Verbatim}[commandchars=@\[\]]
@PYGaE[@# define the eqn f is the C-function defined below]
eqns @PYGbe[=] { @PYGaB[']@PYGaB[z:c]@PYGaB[']: @PYGaB[']@PYGaB[(la+ii*w)*z - Heavi(t-t0)* K*(z-z(t-tau))]@PYGaB['] }
params @PYGbe[=] {
    @PYGaB[']@PYGaB[K]@PYGaB[']  : @PYGaw[0.1] ,
    @PYGaB[']@PYGaB[w]@PYGaB[']  : @PYGaw[1.0],
    @PYGaB[']@PYGaB[la]@PYGaB['] : @PYGaw[0.1],
    @PYGaB[']@PYGaB[tau]@PYGaB[']: pi,
    @PYGaB[']@PYGaB[t0]@PYGaB['] : @PYGaw[2]@PYGbe[*]pi
}

mycode @PYGbe[=] @PYGaB["""]
@PYGaB[double Heavi(double t) {]
@PYGaB[    if(t@textgreater[]=0)]
@PYGaB[        return 1.0;]
@PYGaB[    else]
@PYGaB[        return 0.0;]
@PYGaB[}]
@PYGaB["""]
dde @PYGbe[=] dde23(eqns@PYGbe[=]eqns, params@PYGbe[=]params, supportcode@PYGbe[=]mycode)
\end{Verbatim}

This code would switch a control term on when \code{t\textgreater{}t0}.
Note that \code{Heavi(t-t0)} does not get translated to a delay
term, because \code{Heavi} is not a system variable.

Since this scenario occurs so frequent the Heaviside function (as defined above)
is included by default in the source code.

\subsection{Use and modify generated code}

The compilation of the generated code is done with \code{scipy.weave}.
Instead of using weave to run the code you can directly access the generated
code via the function \code{dde23.output\_ccode()}. This function returns
the generated code as a string which you can then store in a source file.

To run the generated code manually you have to set the precompiler flag\textbackslash{}
\code{\#define MANUAL} (uncomment the line in the source file) to exclude the
python related parts and include some other parts making the code a valid stand
alone source file. After this the code should compile with
\code{g++ -lm -o prog source.cpp} and you can run the program manually.

You can specify the history of all variables in the source file
by setting the \code{for} loops after the comment\textbackslash{}
\code{/* set the history here ...  */}.

Running the code manually can help you debug, if some problem
occurs and also allows you to extend the code easily.

\subsection{Another example}

The following example shows some of the things discussed above.
The code simulates the Lang-Kobayashi laser equations \footnote{
Lang, R.  and Kobayashi, K. , External optical feedback effects on semiconductor injection laser properties, IEEE J. Quantum Electron. 16, 347 (1980)
}
\begin{gather}
\begin{split}E'(t) &= \frac{1}{2}(1+i\alpha) n E + K E(t-\tau)\\
T n'(t) &= p - n - (1+n) | E|^2\end{split}\notag
\end{gather}
\begin{Verbatim}[commandchars=@\[\]]
@PYGal[import] @PYGaW[numpy] @PYGal[as] @PYGaW[np]
@PYGal[import] @PYGaW[pylab] @PYGal[as] @PYGaW[pl]
@PYGal[from] @PYGaW[pydelay] @PYGal[import] dde23

tfinal @PYGbe[=] @PYGaw[10000]
tau @PYGbe[=] @PYGaw[1000]

@PYGaE[@#the laser equations]
eqns @PYGbe[=] {
    @PYGaB[']@PYGaB[E:c]@PYGaB[']: @PYGaB[']@PYGaB[0.5*(1.0+ii*a)*E*n + K*E(t-tau)]@PYGaB['],
    @PYGaB[']@PYGaB[n]@PYGaB[']  : @PYGaB[']@PYGaB[(p - n - (1.0 +n) * pow(abs(E),2))/T]@PYGaB[']
}

params @PYGbe[=] {
    @PYGaB[']@PYGaB[a]@PYGaB[']  : @PYGaw[4.0],
    @PYGaB[']@PYGaB[p]@PYGaB[']  : @PYGaw[1.0],
    @PYGaB[']@PYGaB[T]@PYGaB[']  : @PYGaw[200.0],
    @PYGaB[']@PYGaB[K]@PYGaB[']  : @PYGaw[0.1],
    @PYGaB[']@PYGaB[tau]@PYGaB[']: tau,
    @PYGaB[']@PYGaB[nu]@PYGaB['] : @PYGaw[10]@PYGbe[*]@PYGbe[*]@PYGbe[-]@PYGaw[5],
    @PYGaB[']@PYGaB[n0]@PYGaB['] : @PYGaw[10.0]
}

noise @PYGbe[=] { @PYGaB[']@PYGaB[E]@PYGaB[']: @PYGaB[']@PYGaB[sqrt(0.5*nu*(n+n0)) * (gwn() + ii*gwn())]@PYGaB['] }

dde @PYGbe[=] dde23(eqns@PYGbe[=]eqns, params@PYGbe[=]params, noise@PYGbe[=]noise)
dde@PYGbe[.]set@_sim@_params(tfinal@PYGbe[=]tfinal)

@PYGaE[@# use a dictionary to set the history]
thist @PYGbe[=] np@PYGbe[.]linspace(@PYGaw[0], tau, tfinal)
Ehist @PYGbe[=] np@PYGbe[.]zeros(@PYGaY[len](thist))@PYGbe[+]@PYGaw[1.0]
nhist @PYGbe[=] np@PYGbe[.]zeros(@PYGaY[len](thist))@PYGbe[-]@PYGaw[0.2]
dic @PYGbe[=] {@PYGaB[']@PYGaB[t]@PYGaB['] : thist, @PYGaB[']@PYGaB[E]@PYGaB[']: Ehist, @PYGaB[']@PYGaB[n]@PYGaB[']: nhist}

@PYGaE[@# 'useend' is True by default in hist@_from@_dict and thus the]
@PYGaE[@# time array is shifted correctly]
dde@PYGbe[.]hist@_from@_arrays(dic)

dde@PYGbe[.]run()

t @PYGbe[=] dde@PYGbe[.]sol@PYGZlb[]@PYGaB[']@PYGaB[t]@PYGaB[']@PYGZrb[]
E @PYGbe[=] dde@PYGbe[.]sol@PYGZlb[]@PYGaB[']@PYGaB[E]@PYGaB[']@PYGZrb[]
n @PYGbe[=] dde@PYGbe[.]sol@PYGZlb[]@PYGaB[']@PYGaB[n]@PYGaB[']@PYGZrb[]

spl @PYGbe[=] dde@PYGbe[.]sample(@PYGbe[-]tau, tfinal, @PYGaw[0.1])

pl@PYGbe[.]plot(t@PYGZlb[]:@PYGbe[-]@PYGaw[1]@PYGZrb[], t@PYGZlb[]@PYGaw[1]:@PYGZrb[] @PYGbe[-] t@PYGZlb[]:@PYGbe[-]@PYGaw[1]@PYGZrb[], @PYGaB[']@PYGaB[0.8]@PYGaB['], label@PYGbe[=]@PYGaB[']@PYGaB[step size]@PYGaB['])
pl@PYGbe[.]plot(spl@PYGZlb[]@PYGaB[']@PYGaB[t]@PYGaB[']@PYGZrb[], @PYGaY[abs](spl@PYGZlb[]@PYGaB[']@PYGaB[E]@PYGaB[']@PYGZrb[]), @PYGaB[']@PYGaB[g]@PYGaB['], label@PYGbe[=]@PYGaB[']@PYGaB[sampled solution]@PYGaB['])
pl@PYGbe[.]plot(t, @PYGaY[abs](E), @PYGaB[']@PYGaB[.]@PYGaB['], label@PYGbe[=]@PYGaB[']@PYGaB[calculated points]@PYGaB['])
pl@PYGbe[.]legend()

pl@PYGbe[.]xlabel(@PYGaB[']@PYGaB[@$t@$]@PYGaB['])
pl@PYGbe[.]ylabel(@PYGaB[']@PYGaB[@$@textbar[]E@textbar[]@$]@PYGaB['])

pl@PYGbe[.]xlim((@PYGaw[0.95]@PYGbe[*]tfinal, tfinal))
pl@PYGbe[.]ylim((@PYGaw[0],@PYGaw[3]))
pl@PYGbe[.]show()
\end{Verbatim}

\includegraphics{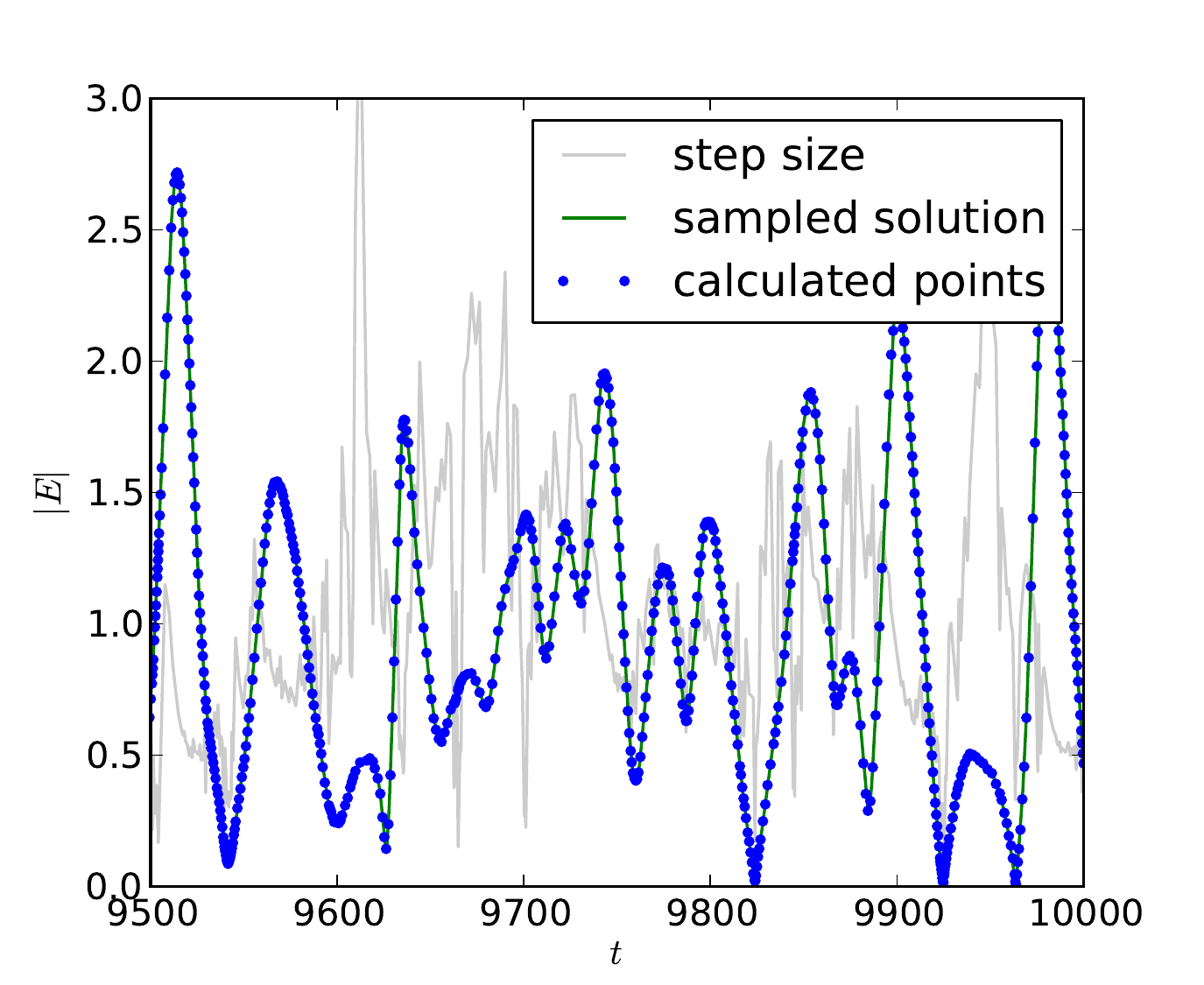}

\section{Module Reference}
\index{\_\_init\_\_() (pydelay.dde23 method)}

\hypertarget{pydelay.dde23.__init__}{}\begin{methoddesc}[dde23]{\_\_init\_\_}{eqns, params=None, noise=None, supportcode='', debug=False}
Initialise the solver.
\begin{description}
\item[\emph{eqns}  ] \leavevmode
Dictionary defining for each variable the derivative. 
Delays are written as as \code{(t-...)}
example:

\begin{Verbatim}[commandchars=@\[\]]
eqns @PYGbe[=] {
    @PYGaB[']@PYGaB[y1]@PYGaB[']: @PYGaB[']@PYGaB[- y1 * y2(t-tau1) + y2(t-tau2)]@PYGaB['],
    @PYGaB[']@PYGaB[y2]@PYGaB[']: @PYGaB[']@PYGaB[a * y1 * y2(t-tau1) - y2]@PYGaB['],
    @PYGaB[']@PYGaB[y3]@PYGaB[']: @PYGaB[']@PYGaB[y2 - y2(t-tau2)]@PYGaB[']
    }
\end{Verbatim}

You can also directly use numbers or combination of parameters as delays:

\begin{Verbatim}[commandchars=@\[\]]
eqns = {
    'x1': '-a*x1 + x1(t - 1.0)',
    'x2': 'x2-b*x1(t-2.0*a+b)
    }
\end{Verbatim}

At the moment only constant delays are supported.

The string defining the equation has to be a valid C expression, i.e.,
use \code{pow(a,b)} instead of \code{a**b} etc. (this might change in the future):

\begin{Verbatim}[commandchars=@\[\]]
eqns @PYGbe[=] {@PYGaB[']@PYGaB[y]@PYGaB[']: @PYGaB[']@PYGaB[-2.0 * sin(t) * pow(y(t-tau), 2)]@PYGaB[']}
\end{Verbatim}

Complex variable can be defined using \code{:C} or \code{:c} in the variable name.
The imaginary unit can be used through \code{ii} in the equations:

\begin{Verbatim}[commandchars=@\[\]]
eqns @PYGbe[=] {@PYGaB[']@PYGaB[z:C]@PYGaB[']: @PYGaB[']@PYGaB[(-la + ii * w0) * z]@PYGaB['] }
\end{Verbatim}

\item[\emph{params} ] \leavevmode
Dictionary defining the parameters (including delays) used in eqns.
example:

\begin{Verbatim}[commandchars=@\[\]]
params @PYGbe[=] {
   @PYGaB[']@PYGaB[a]@PYGaB[']   : @PYGaw[1.0], 
   @PYGaB[']@PYGaB[tau1]@PYGaB[']: @PYGaw[1.0], 
   @PYGaB[']@PYGaB[tau2]@PYGaB[']: @PYGaw[10.0]
   }
\end{Verbatim}

\item[\emph{noise} ] \leavevmode
Dictionary for noise terms. The function \code{gwn()} can be accessed in 
the noise string and provides a Gaussian white noise term of unit variance.
example:

\begin{Verbatim}[commandchars=@\[\]]
noise @PYGbe[=] {@PYGaB[']@PYGaB[x]@PYGaB[']: @PYGaB[']@PYGaB[0.01*gwn()]@PYGaB[']}
\end{Verbatim}

\item[\emph{debug} ] \leavevmode
If set to \code{True} the solver gives verbose output to stdout while running.

\end{description}
\end{methoddesc}
\index{set\_sim\_params() (pydelay.dde23 method)}

\hypertarget{pydelay.dde23.set_sim_params}{}\begin{methoddesc}[dde23]{set\_sim\_params}{tfinal=100, AbsTol=9.9999999999999995e-07, RelTol=0.001, dtmin=9.9999999999999995e-07, dtmax=None, dt0=None, MaxIter=1000000000.0}~\begin{description}
\item[\emph{tfinal} ] \leavevmode
End time of the simulation (the simulation always starts at \code{t=0}).

\item[\emph{AbsTol}, \emph{RelTol} ] \leavevmode
The relative and absolute error tolerance. If the estimated error \emph{e} 
for a variable \emph{y} obeys \emph{e \textless{}= AbsTol + RelTol*\textbar{}y\textbar{}} then the step is accepted.  
Otherwise the step will be repeated with a smaller step size.

\item[\emph{dtmin}, \emph{dtmax} ] \leavevmode
Minimum and maximum step size used.

\item[\emph{dt0} ] \leavevmode
initial step size

\item[\emph{MaxIter} ] \leavevmode
maximum number of steps. The simulation stops if this is reached.

\end{description}
\end{methoddesc}
\index{hist\_from\_arrays() (pydelay.dde23 method)}

\begin{methoddesc}[dde23]{hist\_from\_arrays}{dic, useend=True}
Initialise the history using a dictionary of arrays with variable names as keys.  
Additionally a time array can be given corresponding to the key \code{t}. 
All arrays in \emph{dic} have to have the same lengths.

If an array for \code{t} is given the history is interpreted as points 
\code{(t,var)}. Otherwise the arrays will be evenly spaced out over the interval 
\code{{[}-maxdelay, 0{]}}.

If useend is True the time array is shifted such that the end time is
zero. This is useful if you want to use the result of a previous simulation 
as the history.

If any variable names are missing in the dictionaries, the history of these
variables is set to zero and a warning is printed. 
If the dictionary contains keys not matching any variables (or \code{'t'}) these
entries are ignored and a warning is printed.

Example::

\begin{Verbatim}[commandchars=@\[\]]
t @PYGbe[=] numpy@PYGbe[.]linspace(@PYGaw[0], @PYGaw[1], @PYGaw[500])
x @PYGbe[=] numpy@PYGbe[.]cos(@PYGaw[0.2]@PYGbe[*]t)
y @PYGbe[=] numpy@PYGbe[.]sin(@PYGaw[0.2]@PYGbe[*]t)

histdic @PYGbe[=] {
    @PYGaB[']@PYGaB[t]@PYGaB[']: t,
    @PYGaB[']@PYGaB[x]@PYGaB[']: x,
    @PYGaB[']@PYGaB[y]@PYGaB[']: y
}
dde@PYGbe[.]hist@_from@_arrays(histdic)
\end{Verbatim}
\end{methoddesc}
\index{hist\_from\_funcs() (pydelay.dde23 method)}

\begin{methoddesc}[dde23]{hist\_from\_funcs}{dic, nn=101}
Initialise the histories with the functions stored in the dictionary \emph{dic}.
The keys are the variable names.  The function will be called as \code{f(t)} 
for \code{t} in \code{{[}-maxdelay, 0{]}} on \emph{nn} samples in the interval.

This function provides the simplest way to set the history.
It is often convenient to use python \code{lambda} functions for \code{f}.
This way you can define the history function in place.

If any variable names are missing in the dictionaries, the history of these
variables is set to zero and a warning is printed. If the dictionary contains 
keys not matching any variables these entries are ignored and a warning is 
printed.

Example: Initialise the history of the variables \code{x} and \code{y} with 
\code{cos} and \code{sin} functions using a finer sampling resolution:

\begin{Verbatim}[commandchars=@\[\]]
@PYGal[from] @PYGaW[math] @PYGal[import] sin, cos

histdic @PYGbe[=] {
    @PYGaB[']@PYGaB[x]@PYGaB[']: @PYGay[lambda] t: cos(@PYGaw[0.2]@PYGbe[*]t),
    @PYGaB[']@PYGaB[y]@PYGaB[']: @PYGay[lambda] t: sin(@PYGaw[0.2]@PYGbe[*]t)
}

dde@PYGbe[.]hist@_from@_funcs(histdic, @PYGaw[500])
\end{Verbatim}
\end{methoddesc}
\index{output\_ccode() (pydelay.dde23 method)}

\hypertarget{pydelay.dde23.output_ccode}{}\begin{methoddesc}[dde23]{output\_ccode}{}\end{methoddesc}
\index{run() (pydelay.dde23 method)}

\hypertarget{pydelay.dde23.run}{}\begin{methoddesc}[dde23]{run}{}
run the simulation
\end{methoddesc}
\index{dde23 (class in pydelay)}

\hypertarget{pydelay.dde23}{}\begin{classdesc}{dde23}{eqns, params=None, noise=None, supportcode='', debug=False}
This class translates a DDE to C and solves it using the Bogacki-Shampine method.

\emph{Attributes of class instances:}

\textbf{For user relevant attributes:}
\begin{description}
\item[\emph{self.sol}] \leavevmode
Dictionary storing the solution (when the simulation has finished).
The keys are the variable names and \code{'t'} corresponding to the 
sampled times.

\item[\emph{self.discont} ] \leavevmode
List of discontinuity times. This is generated from the occurring
delays by propagating the discontinuity at \code{t=0}. The solver will step on these 
discontinuities. If you want the solver to step onto certain times they can
be inserted here.

\item[\emph{self.rseed}  ] \leavevmode
Can be set to initialise the random number generator with a specific
seed. If not set it is initialised with the time.

\item[\emph{self.hist}   ] \leavevmode
Dictionary with the history. Don't manipulate the history arrays directly!
Use the provided functions to set the history.

\item[\emph{self.Vhist}  ] \leavevmode
Dictionary with the time derivatives of the history.

\end{description}

\textbf{For user less relevant attributes:}
\begin{description}
\item[\emph{self.delays} ] \leavevmode
List of the delays occurring in the equations.

\item[\emph{self.chunk}  ] \leavevmode
When arrays become to small they are grown by this number.

\item[\emph{self.spline\_tck} ] \leavevmode
Dictionary which stores the tck spline representation of the solutions.
(see \code{scipy.interpolate})

\item[\emph{self.eqns}   ] \leavevmode
Stores the eqn dictionary.

\item[\emph{self.params} ] \leavevmode
Stores the parameter dictionary.

\item[\emph{self.simul}  ] \leavevmode
Dictionary of the simulation parameters.

\item[\emph{self.noise}  ] \leavevmode
Stores the noise dictionary.

\item[\emph{self.debug}  ] \leavevmode
Stores the debug flag.

\item[\emph{self.delayhashs} ] \leavevmode
List of hashs for each delay (this is used in the generated C-code).

\item[\emph{self.vars}    ] \leavevmode
List of variables extracted from the eqn dictionary keys.

\item[\emph{self.types}  ] \leavevmode
Dictionary of C-type names of each variable.

\item[\emph{self.nptypes} ] \leavevmode
Dictionary of numpy-type names of each variable.

\end{description}
\end{classdesc}

\renewcommand{\indexname}{Module Index}

\renewcommand{\indexname}{Index}
\section*{Acknowledgement} We thank Thomas Dahms and Andreas Amann 
for helpful discussions and bug fixing.
\end{document}